\documentclass[11pt, letter]{article}
\usepackage{jheppub}
\usepackage[english]{babel}
\usepackage[dvipsnames]{xcolor}
\usepackage{amsmath,amssymb,slashed,hyperref,mathtools,array,braket,cancel,parskip}
\usepackage[none]{hyphenat} 
\usepackage[none]{hyphenat} 
\usepackage{graphicx}
\usepackage{float}
\usepackage{tikz}
\usepackage{bbold}
\usepackage{comment} 
\usepackage[compat=1.1.0]{tikz-feynman}
\usepackage{todonotes}
\graphicspath{{Pictures/}} 
\newcommand{\nn}{\nonumber}
\newcommand{\sd}{\mathrm{d}}

\newcommand{\bb}[1]{\mathbb{#1}}
\newcommand{\cl}[1]{\mathcal{#1}}

\def\prd{\ref@{Phys.~Rev.~D}}        

\newcommand{\td}[1]{
	\if\notesOn1
	\todo[inline]{#1}
	\fi
}
\hypersetup{
	colorlinks = true,
	linkcolor = Mahogany,
	anchorcolor = Mahogany,
	citecolor = Mahogany,
	filecolor = Mahogany,
	urlcolor = Mahogany
}

\def\notesOn{1}
\tikzset{
	graviton/.style={
		double,
		decoration={snake, aspect=0.75, mirror, segment length=1.5mm},
		decorate
	}
}
\tikzfeynmanset{
	bigblob/.style={
		shape=circle,
		draw=blue,
		fill=red}
}
\DeclareSymbolFont{matha}{OML}{txmi}{m}{it}
\DeclareMathSymbol{\varv}{\mathord}{matha}{118}
\usepackage{enumitem,amssymb}
\newlist{todolist}{itemize}{2}
\setlist[todolist]{label=$\square$}
\usepackage{pifont}
\begin{document}


\title{Kerr-Newman from Minimal Coupling}

\author{Nathan Moynihan}
\affiliation{High Energy Physics, Cosmology \& Astrophysics Theory group\\
	and	The Laboratory for Quantum Gravity \& Strings\\\\
	Department of Mathematics and Applied Mathematics, University of Cape Town, South Africa}

\abstract{
	We show that at 1PN all four-dimensional black hole solutions in asymptotically flat spacetimes can be derived from leading singularities involving minimally coupled three-particle amplitudes. Furthermore, we show that the rotating solutions can be derived from their non-rotating counterparts by a spin-factor deformation of the relevant minimally coupled amplitudes. To show this, we compute the tree-level and one-loop leading singularities for a heavy charged source with generic spin $s$. We compute the metrics both with and without a spin factor and show that we get both the Kerr-Newman and Reissner-Nordstr\"om solutions respectively. We then go on to compute the impulse imparted to the probe particle in the infinite spin limit and show that the spin factor induces a complex deformation of the impact parameter, as was recently observed for Kerr black holes in \cite{Arkani-Hamed:2019ymq}. We interpret these observations as being the on-shell avatar of the Janis-Newman algorithm for charged black holes.
}
\emailAdd{nathantmoynihan@gmail.com}

\maketitle

\section{Introduction}
Extracting classical gravitational physics from quantum field theories has a long history \cite{Thirring:1961dc,Iwasaki:1971vb,Duff:1973zz}. More recently the modern on-shell scattering amplitudes program has provided a number of tools that can be used to greatly simplify calculations of gravitational quantities, notably the KLT relations and the BCJ double copy \cite{Bern:2008qj,Bern:2010ue,Bern:2010yg,Johansson:2019dnu,Bautista:2019evw}, as well as those related specifically to classical observables \cite{Guevara:2017csg,Kosower:2018adc,Maybee:2019jus}. While the original aim of the double copy program was to simplify loop computations in gravity, it has found many uses in classical gravity, from metric reconstruction \cite{Neill:2013wsa,Luna:2016hge,Monteiro:2014cda,Luna:2015paa,Carrillo-Gonzalez:2017iyj,Luna:2018dpt,CarrilloGonzalez:2019gof,LopesCardoso:2018xes} to gravitational wave physics \cite{Goldberger:2016iau,Luna:2017dtq,Bern:2019crd,Bautista:2019tdr}. In particular, the introduction of a formalism to compute amplitudes of arbitrary mass and spin \cite{Arkani-Hamed:2017jhn} has provided a powerful way to investigate spin effects in classical observables \cite{Vaidya:2014kza,Chung:2018kqs,Guevara:2018wpp,Bautista:2019tdr,Guevara:2019fsj}. Calculations involving spin effects in gravity are often computed in the post-Newtonian (small velocities $v\ll c$) or post-Minkowskian (expansion in $G$) frameworks \cite{Porto:2005ac,Levi:2010zu,Levi:2014gsa,Levi:2015msa,Damour:2016gwp,Vines:2016qwa,Vines:2017hyw,Vines:2018gqi,Levi:2018nxp}, however there have also been calculations involving loop amplitudes via standard Feynman diagram techniques and form factors \cite{Holstein:2008sy,Holstein:2008sx}. Moreover, recent work by a number of authors have shown that such calculations can be efficiently streamlined by using modern amplitude techniques, often combined with the tools of effective field theory \cite{Bjerrum-Bohr:2013bxa,Damour:2016gwp,Cheung:2018wkq,Bjerrum-Bohr:2018xdl,Carballo-Rubio:2018bmu,Cristofoli:2019neg,Bern:2019nnu}.

In four dimensions, black holes are classically described only by their mass, angular momentum and charge by the no hair theorem. In particular, the unique stationary, asymptotically flat black hole with all of these properties (with non-degenerate horizons) is the Kerr-Newman black hole \cite{Mazur:1982db}, making it the most general black hole in our universe. From far enough away, any black hole can be treated as a point particle, and as such can be given an effective one-body description. The proposed on-shell avatar of the no-hair theorem is that black hole solutions should be obtainable from \textit{minimal coupling}, with deviations describing finite-size effects given by non-minimal deformations \cite{Chung:2018kqs,Arkani-Hamed:2019ymq}. The construction of classical and quantum black hole metrics using loop amplitudes has been a fruitful endeavour, using everything from form factors \cite{Donoghue:2001qc,BjerrumBohr:2002ks} to unitarity based methods \cite{Neill:2013wsa} and more recently with leading singularities \cite{Chung:2018kqs,Emond:2019crr,Chung:2019duq}. In this paper, we show that all four-dimensional black hole solutions at order $G$ and charge $\alpha$ are obtainable from minimal coupling via the tree-level and one-loop triangle leading singularities. Furthermore, we show explicitly that the relevant amplitudes themselves factorise into a spin-independent piece and a spin factor, as was demonstrated in the case of Kerr black holes in Refs. \cite{Chung:2018kqs,Arkani-Hamed:2019ymq,Chung:2019duq}. Very recently, it was shown that this factorization, in the infinite spin limit, is the on-shell avatar of the Janis-Newman algorithm \cite{Arkani-Hamed:2019ymq}, which utilises a complex coordinate transformation of the Schwarzchild (Riessner-Nordstr\"om) solution leading directly to the Kerr (Kerr-Newman) solution \cite{Newman:1965tw,Newman:1965my}. We will show that the Kerr-Newman solution can be derived in precisely this way from Reissner-Nordstr\"om by simply attaching a spin-factor to the relevant minimally coupled three-point amplitudes.

We will consider a scalar test particle $p_1$ gravitationally probing a heavy, charged, spinning source with momentum $p_3$. We will take particles $p_1, p_2$ to be massive particles with mass $m_A$, and particles $p_3, p_4$ to be spinning with mass $m_B$ and spin $s$.
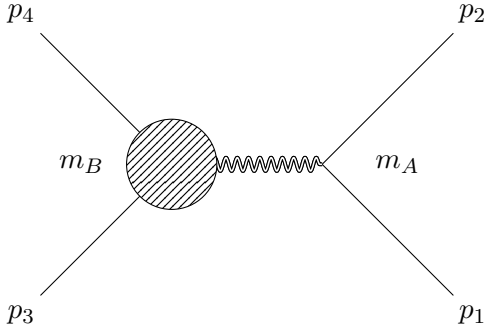
\begin{figure}[H]
	\centering
	\begin{tikzpicture}[scale=1]
	\begin{feynman}  
	\vertex (a) at (-4,2) {$p_4$};
	\vertex (b) at (-4,-2) {$p_3$};
	\vertex (c) at (2,-2) {$p_1$};
	\vertex (d) at (2,2) {$p_2$};
	\vertex (r) at (0,0);
	\vertex (l) at (-2,0) ;
	\diagram* {
		(a) -- [plain] (l) -- [graviton] (r) -- [plain] (d),
		(b) -- [plain] (l) -- [graviton] (r) -- [plain] (c),
	};
	\draw[preaction={fill, white},pattern=north east lines] (-2,0) ellipse (0.6cm and 0.6cm);
	\draw (-3.2,-0.25) node[above] {$m_B$};
	\draw (1,-0.25) node[above] {$m_A$};
	\end{feynman}
	\end{tikzpicture}
	\caption{Gravitational probe of charged, spinning particles}
	\label{triangleLS}
\end{figure}
\section{Scattering Amplitudes and Spin Operators}
In the textbook formulation of quantum field theory, the familiar Gordon decomposition identity is given by
\begin{equation}\label{gordonidentity}
\bar{u}(p_1)\gamma^\mu u(p_2) = \bar{u}(p_1)\left[\frac{p_1^\mu + p_2^\mu}{2m} + \frac{i\sigma^{\mu\nu}(p_1^\mu - p_2^\mu)}{2m}\right]u(p_2).
\end{equation} 
This identity has many uses, e.g. expressing the vertex function of a massless photon interacting with two massive fermions in terms of form factors, one corresponding to spin-independent and spin-dependent parts. 
In \cite{Arkani-Hamed:2017jhn}, it was shown that the on-shell avatar of this identity is that one can expose the spin-dependence of an on-shell `vertex' by choosing a purely chiral spinor basis. 

Suppose we want to express everything in an anti-chiral basis. Using the formalism of \cite{Arkani-Hamed:2017jhn}, we find that the three particle amplitude in the undotted frame is given by 
\begin{equation}\label{key}
\cl{M}_{\bar{f}f\gamma} = \bar{u}_1\cancel{\epsilon_3}v_2 = x\varepsilon^{\alpha_1\alpha_2}.
\end{equation}
Converting between bases is done with the operator $p/m$, which means that for this amplitude in the dotted frame we find
\begin{equation}\label{key}
\cl{M}_{\bar{f}f\gamma} = x\varepsilon^{\alpha_1\alpha_2} = x\frac{\varepsilon^{\alpha_1\alpha_2}p_{1\alpha_1\dot{\alpha}_1}p_{2\alpha_2\dot{\alpha}_2}}{m^2} = x\varepsilon_{\dot{\alpha}_1\dot{\alpha}_2} + \frac{\tilde{\lambda}_{3\dot{\alpha}_1}\tilde{\lambda}_{3\dot{\alpha}_2}}{m},
\end{equation}
where we have used the identities
\begin{equation}\label{key}
\cl{O}_{\alpha\beta} \coloneqq \frac{p_{1\alpha}^{~~\dot{\alpha}}p_{2\beta\dot{\alpha}}}{m^2} = \varepsilon_{\alpha\beta} -x\frac{\lambda_{3\alpha}\lambda_{3\beta}}{m},~~~~~\cl{O}_{\dot{\alpha}\dot{\beta}} \coloneqq \varepsilon_{\dot{\alpha}\dot{\beta}} + \frac{1}{x}\frac{\tilde{\lambda}_{3\dot{\alpha}_1}\tilde{\lambda}_{3\dot{\alpha}_2}}{m}.
\end{equation}

To see how this relates to the spin, we consider the Pauli-Lubanski pseudo-vector $S^\mu = -\frac{1}{2m}\epsilon^{\mu\nu\rho\sigma}p_{\nu}\sigma_{\rho\sigma}$, where
\begin{equation}\label{key}
(\sigma_{\mu\nu})_{\alpha}^{~~\beta} = \frac{i}{2}(\sigma_{[\mu}\bar{\sigma}_{\nu]})_{\alpha}^{~~\beta},~~~~~(\bar{\sigma}_{\mu\nu})^{\dot{\alpha}}_{~~\dot{\beta}} = -\frac{i}{2}(\bar{\sigma}_{[\mu}\sigma_{\nu]})^{\dot{\alpha}}_{~~\dot{\beta}}.
\end{equation}
For chiral $SL(2,\mathbb{C})$ representations of massive states, we can write a general spin-$s$ generator $\bar{\sigma}_{\mu\nu}$ in a simpler form, due to the fact that the external polarization tensors are always built from symmetrized massive spinors, meaning we can write
\begin{equation}\label{key}
(\bar{\sigma}_{\mu\nu})^{\dot{\alpha}_1\cdots \dot{\alpha}_{2s}}_{~~~~\dot{\beta}_1\cdots\dot{\beta}_{2s}} = \sum_i(\sigma_{\mu\nu})^{\dot{\alpha}_i}_{~~\dot{\beta}_i}\bar{\bb{I}}_i,
\end{equation}
where $\bar{\bb{I}}_i =  \delta^{\dot{\alpha}_1}_{\dot{\beta}_1}\cdots\delta^{\dot{\alpha}_{i-1}}_{\dot{\beta}_{i-1}}\delta^{\dot{\alpha}_{i+1}}_{\dot{\beta}_{i+1}}\cdots \delta^{\dot{\alpha}_{2s}}_{\dot{\beta}_{2s}}$, with $\sigma_{\mu\nu}$ and $\bb{I}_i$ given analogously. We can therefore write
\begin{align}\label{key}
(S_\mu)^{\dot{\alpha}}_{~~\dot{\beta}} &= \frac{i}{m}p_\nu (\bar{\sigma}^{\mu\nu})^{\dot{\alpha}}_{~~\dot{\beta}}\nonumber\\
&= \frac{1}{4m}\left[(p\cdot\sigma)\bar{\sigma}_\mu - \sigma_\mu (p\cdot\bar{\sigma})\right]^{\dot{\alpha}}_{~~\dot{\beta}},
\end{align}
where we have used the identity $\sigma^{\mu\nu} = -\frac{i}{2}\epsilon^{\mu\nu\rho\sigma}\sigma_{\rho\sigma}$. We can generalise this for \textit{any} spin $s$ by noting that $\sum_i(\bar{\sigma}_{\mu\nu})^{\dot{\alpha}_i}_{~~\dot{\beta}_i}\bar{\bb{I}}_i = 2s(\bar{\sigma}_{\mu\nu})^{\dot{\alpha}_1}_{~~\dot{\beta}_1}\bar{\bb{I}}_1$ to find, in spinor helicity notation,
\begin{equation}\label{key}
(S_\mu)^{\dot{\alpha}_1\cdots \dot{\alpha}_{2s}}_{~~~~\dot{\beta}_1\cdots\dot{\beta}_{2s}} = \frac{s}{2m}\left(\bra{\textbf{p}}\sigma_\mu|\textbf{p}] + [\textbf{p}|\bar{\sigma}_\mu\ket{\textbf{p}}\right)\bar{\bb{I}}_1.
\end{equation}
Contracting this with an external massless momentum $p_3$, we then find
\begin{align}\label{key}
(p_3\cdot S)^{\dot{\alpha}}_{~~\dot{\beta}} &= -\frac{|3][3|}{2x}
\end{align}
where we have used $(p\cdot\sigma)_{\alpha\dot{\alpha}} = -\ket{p}_{\alpha}[p|_{\dot{\alpha}}$ and $(p\cdot\bar{\sigma})^{\dot{\alpha}\alpha} = -|p]^{\dot{\alpha}}\bra{p}^{\alpha}$.

We can now establish the spin-dependence of a three particle amplitude with two spinning particles coupled to a massless particle of (positive) helicity $h$
\begin{equation}\label{key}
\cl{M}_3^{s,h} = g(mx)^h\frac{\braket{\textbf{12}}^{2s}}{m^{2s}} = -g(mx)^h\left[[\textbf{1}|\left(1 - \frac{|3][3]}{mx}\right)|\textbf{2}]\right]^{2s}.
\end{equation}
We will be interested in computing leading singularities (LS) throughout the rest of this paper and as such we will strip off the external spinors, expressing amplitudes in a basis of un-contracted purely anti-chiral indices, as is suited for LS calculations \cite{Cachazo:2017jef,Guevara:2017csg}. While there can be additional spin-dependence coming from these external wavefunctions, we will see that these can be restored after the fact by considering the contributions from the non-chiral or `polarization tensor' basis.
\section{Tree-Level Leading Singularity}
At tree level, there is only one possible diagram that we can consider
\begin{figure}[H]
	\centering
	\begin{tikzpicture}[scale=1]
	\begin{feynman}  
	\vertex (a) at (-4,2) {$p_4$};
	\vertex (b) at (-4,-2) {$p_3$};
	\vertex (c) at (2,-2) {$p_1$};
	\vertex (d) at (2,2) {$p_2$};
	\vertex (r) at (0,0);
	\vertex (l) at (-2,0) ;
	\diagram* {
		(a) -- [plain] (l) -- [graviton] (r) -- [plain] (d),
		(b) -- [plain] (l) -- [graviton] (r) -- [plain] (c),
	};
	\draw (-3.2,-0.25) node[above] {$m_B$};
	\draw (1,-0.25) node[above] {$m_A$};
	\end{feynman}
	\end{tikzpicture}
	\caption{Tree Level Diagram}
	\label{treediag}
\end{figure}
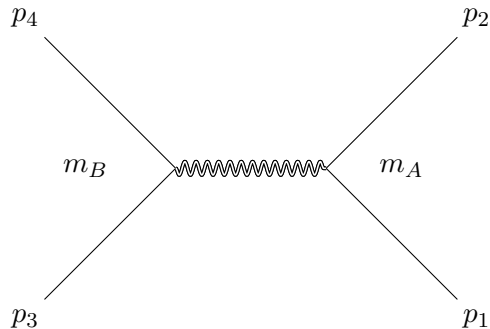
Since there is no electromagnetic interaction, this will simply produce a purely gravitational interaction at order $G$, and has been calculated many times in the literature \cite{Arkani-Hamed:2019ymq,Guevara:2018wpp,Chung:2018kqs}. However for completeness, and in order to set notation, we will briefly review the calculation of this piece here.

In this paper, we are only going to concern ourselves with the classical effects, and as such to greatly simplify calculations we will appeal to the \textit{Holomorphic Classical Limit} (HCL) \cite{Guevara:2017csg}. This allows us to parametrise our scattering amplitudes in such a way that the classical limit can be taken cleanly. While we will not require much of the technical machinery of the HCL in this section, we will implicitly drop any terms that don't survive in the HCL.

The minimal coupling two spin-$s$ one graviton amplitude is given by
\begin{equation}\label{key}
\cl{M}_3[1,2,K^{+2}] = \frac{\kappa}{2}(mx_{12})^2\frac{\braket{\textbf{12}}^{2s}}{m^{2s}},~~~~~\cl{M}_3[1,2,K^{-2}] = \frac{\kappa}{2}\left(\frac{m}{x_{12}}\right)^2\frac{[\textbf{12}]^{2s}}{m^{2s}}
\end{equation}   

Stripping off the external wavefunctions and bearing in mind the discussion in the last section, we can rewrite the three-point amplitudes as
\begin{equation}\label{key}
\cl{M}_L[1,2,K^{+2}] = \frac{\kappa}{2}(mx_{12})^{2}\left(\bb{1} + \frac{K\cdot a}{s}\right)^{2s},~~~~~\cl{M}_L[1,2,K^{-2}] = \frac{\kappa}{2}\left(\frac{m}{x_{12}}\right)^2\left(\bb{1} - \frac{K\cdot a}{s}\right)^{2s},
\end{equation}
where we have defined the anti-chiral, spin-$s$ mass-rescaled Pauli--Lubanski pseudovectors as\footnote{We note that the \textit{classical} value of the spin is obtained by taking $s\rightarrow\infty$ while keeping $s\hbar$ fixed, meaning we take the spin vector to contain a factor of $2s\hbar$.}
\begin{equation}\label{key}
(a_{i,\mu})^{\dot{\alpha}_1\cdots \dot{\alpha}_{2s}}_{~~\dot{\beta}_1\cdots\dot{\beta}_{2s}} = -\frac{2is}{m_i^2}(P^\nu_i \bar{\sigma}_{\mu\nu})^{\dot{\alpha}_i}_{~~\dot{\beta}_i}\bar{\bb{I}}_i,
\end{equation}
where $\bar{\bb{I}}_i = \delta^{\dot{\alpha}_1}_{\dot{\beta}_1}\cdots\delta^{\dot{\alpha}_{i-1}}_{\dot{\beta}_{i-1}}\delta^{\dot{\alpha}_{i+1}}_{\dot{\beta}_{i+1}}\cdots \delta^{\dot{\alpha}_{2s}}_{\dot{\beta}_{2s}}$ 
and where $x_{ij}$ is defined via 
\begin{equation}\label{key}
x_{ij}\lambda_i^\alpha = \frac{\tilde{\lambda}_{i\dot{\alpha}}P^{\dot{\alpha}\alpha}_j}{m},~~~~~\frac{\tilde{\lambda}_i^{\dot{\alpha}}}{x_{ij}} = \frac{p_j^{\dot{\alpha}\alpha}\lambda_{i\alpha}}{m}.
\end{equation}
Choosing to work in the anti-chiral basis means we only consider the spin factor of positive helicity amplitudes. Making this choice, we can glue together two three-points in the $t$ channel to find
\begin{align}\label{key}
\cl{M}_4[1,2,3^{s},4^{s}] &= \left(\frac{\kappa}{2}\right)^2\frac{m_A^2m_B^2}{t}\left(\frac{x_{34}^2}{x_{12}^2}\left(\bb{1} + \frac{K\cdot a}{s}\right)^{2s} + \frac{x_{12}^2}{x_{34}^2}\right),
\end{align}
where any other pieces that contribute to the amplitude vanish in the HCL. We note that we have stripped off the Kronecker deltas that carry the explicit anti-chiral indices, following the conventions of the HCL set out in \cite{Guevara:2017csg}.

We now define the variables
\begin{equation}\label{key}
u = m_Am_B\frac{x_{34}}{x_{12}},~~~~~v = m_Am_B\frac{x_{12}}{x_{34}}.
\end{equation}
Using these definitions, we can derive the following useful identities
\begin{equation}\label{key}
uv = m_A^2m_B^2,~~~~~u + v = 2p_1\cdot p_3.
\end{equation}
We can use this system of equations to derive the individual expressions for $u$ and $v$
\begin{align}\label{eq:uvmand}
2u &= s-m_{A}^{2}-m_{B}^{2}+\sqrt{\left((m_A-m_B)^2 - s\right)\left((m_A+m_B)^2 - s\right)}\nn\\&= 2m_Am_B(\rho + \sqrt{\rho^2-1})\\
2v &= s-m_{A}^{2}-m_{B}^{2}-\sqrt{\left((m_A-m_B)^2 - s\right)\left((m_A+m_B)^2 - s\right)}\nn\\&= 2m_Am_B(\rho - \sqrt{\rho^2-1}),
\end{align}
where we have defined $\rho \coloneqq \frac{p_1\cdot p_3}{m_Am_B}$. With this notation, taking the non-relativistic limit coincides with taking $\rho\longrightarrow 1$. However, taking this limit naively typically obscures the spin dependence, and hence we will need to consider higher orders, expanded around $\rho = 1$. In the classical potential, the spin dependence is expected to show up as \cite{Holstein:2008sx}
\begin{equation}\label{key}
\epsilon_{\mu\nu\rho\sigma}p_1^\mu p_3^\nu K^\rho S^\sigma = m_B(E_A + E_B)(\textbf{a}\cdot \textbf{p}\times \textbf{q}).
\end{equation}
We note that here in the centre of mass frame, $\textbf{p}_1 = -\textbf{p}_3 = \textbf{p} + \frac12\textbf{q}$, meaning we can freely exchange $\textbf{p}$ with $\textbf{p}_1$ or $-\textbf{p}_3$ in the above expression.

Expressing the Gram determinant above in terms of more familiar variables, we find
\begin{align}\label{spinmatching}
i\epsilon_{\mu\nu\rho\sigma}p_1^\mu p_3^\nu K^\rho S^\sigma &= \frac12K\cdot S\sqrt{\left((m_A-m_B)^2 - s\right)\left((m_A+m_B)^2 - s\right)}\nn\\
&= m_Am_B\sqrt{\rho^2 - 1}K\cdot S.
\end{align}
Thus, as promised, we will need to keep up to at least $\cl{O}(\sqrt{\rho^2 - 1})$ in the expansion of $u,v$ and make the above identification before taking the $\rho\longrightarrow 1$ limit. The strategy to obtain spin-dependence at all orders is to expand $u,v$ in powers of $\sqrt{\rho^2 -1}$, matching order by order with eq. \ref{spinmatching}.

With this in mind, the tree-level leading singularity is
\begin{align}\label{kerr4pt}
\cl{M}_4^s &= \left(\frac{\kappa}{2}\right)^2\frac{1}{t}\left(u^2\left(\bb{1} + \frac{K\cdot a}{s}\right)^{2s} + v^2\right),
\end{align}
where we have defined $a = 2s\tilde{a}$.
\section{One-Loop Leading Singularity}
So far we have only considered a purely gravitational interaction, but ultimately we wish to consider black holes that carry charge, and as such we require there to be both a gravitational and an electromagnetic interaction between the scattered objects. There is no tree-level scattering amplitude that can achieve this, and so we must consider at minimum a one-loop process. Since we are are not interested in quantum effects at this stage, we consider only the triangle leading singularity (LS) which is expected to give us a multiple discontinuity in the $t$-channel leading to classical effects \cite{Cachazo:2017jef}.
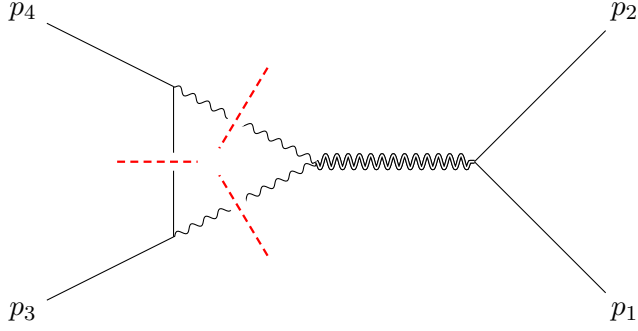
\begin{figure}[H]
	\centering
	\begin{tikzpicture}[scale=1]
	\begin{feynman}
	\vertex (m) at ( -2, 1);
	\vertex (mp) at ( -2, -1);
	\vertex (q) at (2, 0);
	\vertex (qp) at (2,0);
	\vertex (x) at (-0.1,0);
	\vertex (xp) at (0.1,0);
	\vertex (y) at (-0.1,0);
	\vertex (yp) at (0.1,0);  
	\vertex (a) at (-4,2) {$p_4$};
	\vertex (b) at (-4,-2) {$p_3$};
	\vertex (c) at (4,-2) {$p_1$};
	\vertex (d) at (4,2) {$p_2$};
	\diagram* {
		(a) -- [plain] (m) -- [photon] (x) -- [graviton] (q) -- [plain] (d),
		(m) -- [plain] (mp),
		(b) -- [plain] (mp) -- [photon] (x) -- [graviton] (qp)  -- [plain] (c)
	};
	\fill[white] (-2,0) ellipse (0.12cm and 0.12cm);
	\fill[white] (-1.15,0.58) ellipse (0.13cm and 0.13cm);
	\fill[white] (-1.165,-0.54) ellipse (0.13cm and 0.13cm);
	\draw [densely dashed, red, line width=0.3mm,] (-0.75,1.25) -- (-1.5,0);
	\draw [densely dashed, red, line width=0.3mm,] (-0.75,-1.25) -- (-1.5,0);
	\draw [densely dashed, red, line width=0.3mm,] (-2.75,0) -- (-1.5,0);
	\fill[white] (-1.48,0) ellipse (0.2cm and 0.2cm);
	\end{feynman}
	\end{tikzpicture}
	\caption{LS Triangle Diagram}
	\label{triangleLS}
\end{figure}
The only diagram we need consider is the one in Fig. \ref{triangleLS}, where the two massless exchange particles are photons with opposite helicity\footnote{Two same-helicity photons do not contribute to the LS as they have zero residue.} coupled to a graviton. The LS is then given by
\begin{equation}\label{key}
\cl{I} = \sum_{h=\pm}\oint_{\Gamma} \frac{\sd^4L}{(L^2-m^2)k_3^2k_4^2}\cl{M}_3[p_3^s,-L,k_3^{h}]\cl{M}_3[L,p_4^s,k_4^{-h}]\cl{M}_4[-k_3^{-h},-k_4^{h},p_1,p_2],
\end{equation}
where $k_3 = -L + p_3$ and $k_4 = L - p_4$.

We define the exchanged momentum as
\begin{equation}\label{key}
K = |\lambda]\bra{\lambda} = (0,\textbf{q}),~~~~~K^2 = t = -|\textbf{q}|^2,
\end{equation}
which, along with the results and notation from \cite{Arkani-Hamed:2017jhn,Guevara:2017csg,Burger:2017yod}, allow us to express the required tree-level amplitudes as
\begin{equation}\label{key}
\cl{M}_3[1^s,2^s,K^{+1}] = \sqrt{2}emx_{12}\left(\bb{1} + \frac{K\cdot a}{s}\right)^{2s},~~~~~\cl{M}_3[1^s,2^s,K^{-1}] = \sqrt{2}e\left(\frac{m}{x_{12}}\right),
\end{equation}
\begin{equation}\label{key}
\cl{M}_4[k_3^{-1},k_4^{+1},1,2] = -\left(\frac{\kappa}{2}\right)^2\left(m^2\frac{x_{k_3p_2}}{x_{k_4p_1}}\right) = -\left(\frac{\kappa}{2}\right)^2\left(m^2\frac{x_{k_3p_1}}{x_{k_4p_2}}\right)
\end{equation}
\begin{equation}\label{key}
\cl{M}_4[k_3^{+1},k_4^{-1},1,2] = -\left(\frac{\kappa}{2}\right)^2\left(m^2\frac{x_{k_4p_2}}{x_{k_3p_1}}\right) = -\left(\frac{\kappa}{2}\right)^2\left(m^2\frac{x_{k_4p_1}}{x_{k_3p_2}}\right)
\end{equation} 
To make this problem tractable, we work in a parametrisation that makes the classical pieces explicit, e.g. the one given in \cite{Guevara:2017csg}:
\begin{equation}
\begin{split}
p_{3} & =|\eta]\langle\lambda|+|\lambda]\langle\eta|\,,\\
p_{4} & =\beta|\eta]\langle\lambda|+\frac{1}{\beta}|\lambda]\langle\eta|+|\lambda]\langle\lambda|\,,\\
\frac{t}{m_{b}^{2}} & =\frac{(\beta-1)^{2}}{\beta}\,,\\
\langle\lambda\eta\rangle & =[\lambda\eta]=m_{B}\,.
\end{split}
\label{eq:param}
\end{equation}
In addition, we parametrise the loop momentum $L$ as
\begin{equation}\label{key}
L = z\ell + \omega K,~~~~~|\ell] = |\eta] + B|\lambda],~~~~~\bra{\ell} = \bra{\eta} + A\bra{\lambda}.
\end{equation}
Demanding the on-shell cut conditions $k_{3,4}^2 = L^2 - m_B^2$ fixes $\omega = -\frac{1}{z}$ with $A = -B = -\frac{1}{z}\frac{2\beta}{1+\beta}$. This fixes the integration to become
\begin{equation}\label{key}
\frac{\beta}{8(\beta^2 - 1)m_B^2}\oint_{\Gamma}\frac{\sd y}{y} = \frac{1}{16\sqrt{-t}m_B}\oint_{\Gamma}\frac{\sd y}{y},
\end{equation}
where we have taken the $\beta\longrightarrow 1$ limit.

The chosen parameters also induces a convenient parametrisation for $k_{3,4}$
\begin{align}\label{key}
|k_3] &= \frac{1}{\beta+1}\left(|\eta](\beta^{2}-1)y+|\lambda](1+\beta y)\right),\nonumber\\
\bra{k_3} &= \frac{1}{\beta+1}\left(\langle\eta|(\beta^{2}-1)-\frac{1}{y}\langle\lambda|(1+\beta y)\right),\nonumber\\
|k_4] &= \frac{1}{\beta+1}\left(-\beta|\eta](\beta^{2}-1)y+|\lambda](1-\beta^{2}y)\right),\nonumber\\
\bra{k_4} &= \frac{1}{\beta+1}\left(\frac{1}{\beta}\langle\eta|(\beta^{2}-1)+\frac{1}{y}\langle\lambda|(1-y)\right).
\end{align}
When required, we can also evaluate these directly in the HCL $\beta\rightarrow 1$, finding
\begin{equation}\label{key}
|k_3] = \frac12 |\lambda](1+y),~~~~~\bra{k_3} = \frac{1}{2y}\bra{\lambda}(1+y),~~~~~|k_4] = \frac12 |\lambda](1-y),~~~~~\bra{k_4} = -\frac{1}{2y}\bra{\lambda}(1-y).
\end{equation}
In order to perform the contour integral we need to make all factors of $y$ explicit. Conveniantly, in this parametrisation, we find that
\begin{equation}\label{key}
x_{k_ip_{j}} = -y\frac{\bra{\lambda}p_{j}|\eta]}{m_jm_B},~~~~~\frac{1}{x_{k_ip_{j}}} = -\frac{1}{y}\frac{\bra{\eta}p_{j}|\lambda]}{m_jm_B},
\end{equation}
meaning that $x_{k_3p_3} = x_{k_4p_4} = -y$. 

With this set of parameters in place, we can express the product of three particle amplitudes as
\begin{equation}\label{key}
\cl{M}_3[p_3^s,-L^s,-k_3^{\pm 1}]\cl{M}_3[-p_4^s,L^s,-k_4^{\mp 1}] = 2e^2m_B^2\left(1 \pm \frac{(1\pm y)^2}{2y}K\cdot a\right)^{2s}.
\end{equation}
The four particle amplitude is given by
\begin{align}\label{fourpt}
\cl{M}_4[k_3^{-1},k_4^{+1},p_1,p_2] &= -\left(\frac{\kappa}{2}\right)^2\frac{\bra{k_4}p_1|k_3]^2}{t}\nn\\
&= -\left(\frac{\kappa}{2}\right)^2\left(\frac{\left(1-y^2\right) (v-u)}{4 y}+\frac{1}{2} u (1-y)+\frac{1}{2} v (y+1)\right)^2\\
&\simeq -\left(\frac{\kappa}{2}\right)^2m_A^2\left(1 - \frac{\epsilon(1+y)^2}{2y}\right)^2,\nn
\end{align}
where $\epsilon = \sqrt{\rho^2 - 1}$.

We find the LS that we need to evaluate is then
\begin{align}\label{LS2}
\cl{I}_{s} &= \frac{gm_A^2m_B}{16\sqrt{-t}}\oint_{\Gamma} \frac{\sd y}{y} \left[ \left(1-\frac{\epsilon(1+y)^2}{2y}\right)^2\left(1+\frac{(1+y)^2}{4y}\frac{K\cdot a}{s}\right)^{2s}\right],
\end{align}
where the sign difference that would come from the spin factor being attached to the opposite vertex is account for by evaluating the residue at both $y=0$ and $y=\infty$. This explicit form makes it obvious that one has to evaluate $u,v$ beyond the simple non-relativistic limit (for finite spin $s$) in order to observe spin effects, as discussed previously.
\section{Classical Potential}
Now that we have computed the order $\cl{O}(G)$ and order $\cl{O}(e^2)$ leading singularities, we can proceed to compute the classical potential from the HCL. This will allow us to compute the spin-dependent parts of the potential from the sum of the two LS's, diagrammatically given by
\begin{figure}[H]
\centering
\begin{equation}
\begin{gathered}
\begin{tikzpicture}[scale=0.6]
\begin{feynman}  
\vertex (a) at (-4,2) {$p_4$};
\vertex (b) at (-4,-2) {$p_3$};
\vertex (c) at (2,-2) {$p_1$};
\vertex (d) at (2,2) {$p_2$};
\vertex (r) at (0,0);
\vertex (l) at (-2,0) ;
\diagram* {
	(a) -- [plain] (l) -- [graviton] (r) -- [plain] (d),
	(b) -- [plain] (l) -- [graviton] (r) -- [plain] (c),
};
\draw[preaction={fill, white},pattern=north east lines] (-2,0) ellipse (0.6cm and 0.6cm);
\end{feynman}
\end{tikzpicture}
\end{gathered}
=
\begin{gathered}
\begin{tikzpicture}[scale=0.6]
\begin{feynman}  
\vertex (a) at (-4,2) {$p_4$};
\vertex (b) at (-4,-2) {$p_3$};
\vertex (c) at (2,-2) {$p_1$};
\vertex (d) at (2,2) {$p_2$};
\vertex (r) at (0,0);
\vertex (l) at (-2,0) ;
\diagram* {
	(a) -- [plain] (l) -- [graviton] (r) -- [plain] (d),
	(b) -- [plain] (l) -- [graviton] (r) -- [plain] (c),
};
\end{feynman}
\end{tikzpicture}
\end{gathered}
+
\begin{gathered}
	\begin{tikzpicture}[scale=0.6]
\begin{feynman}
\vertex (m) at ( -2, 1);
\vertex (mp) at ( -2, -1);
\vertex (q) at (2, 0);
\vertex (qp) at (2,0);
\vertex (x) at (-0.1,0);
\vertex (xp) at (0.1,0);
\vertex (y) at (-0.1,0);
\vertex (yp) at (0.1,0);  
\vertex (a) at (-4,2) {$p_4$};
\vertex (b) at (-4,-2) {$p_3$};
\vertex (c) at (4,-2) {$p_1$};
\vertex (d) at (4,2) {$p_2$};
\diagram* {
	(a) -- [plain] (m) -- [photon] (x) -- [graviton] (q) -- [plain] (d),
	(m) -- [plain] (mp),
	(b) -- [plain] (mp) -- [photon] (x) -- [graviton] (qp)  -- [plain] (c)
};
\end{feynman}
\end{tikzpicture}
\end{gathered}
\end{equation}
\caption{Diagrams contributing to the classical potential at order $G$ and $\alpha$.}
\label{triangleLS}
\end{figure}
At this point, it is pertinent to explain how working in the chiral basis obscures certain factors that would be observed otherwise, e.g. if we were to work in the non-chiral (polarisation) basis. In Ref. \cite{Guevara:2018wpp} it was proposed that these additional terms could be exposed by considering the \textit{Generalised Expectation Value} (GEV), which amounts to normalising the LS in such a way that the information is restored. It was shown that the normalisation that one needs to take into account is given by the product of massive polarization tensors of the external particles. 
For our purposes, since we have stripped external spinors, we will simply use the perturbative exponential normalisation given in \cite{Guevara:2018wpp}, namely that we need to include a factor of $e^{-K\cdot a}$ for each positive helicity particle. Purturbatively expanding this exponential (for small transfer momentum $K$) to the required order and matching with $\sqrt{\rho^2 - 1}$ to determine the spin contributions will restore the information obscured by working in the purely chiral basis. We note that we drop all terms not linear in $K\cdot a$ after the spin identification has been made. This was also shown in \cite{Chung:2019duq} as being the factor that one picks up when comparing the residue calculated in the polarization tensor basis with one in the anti-chiral basis. Furthermore, we note that an additional spin-dependent term can be picked up from the product of polarisation tensor contractions that we are missing working in an unpolarised expansion. This was calculated in \cite{Chung:2018kqs} and found, to first order, to be
\begin{equation}\label{polcontribution}
\epsilon^\star(p_3)\epsilon(p_4) = \epsilon^\star(p)\left[\bb{1} - \frac{i}{2m_B}\left(\textbf{a}\cdot(\textbf{p}\times \textbf{q})\right)\right]\epsilon(p),
\end{equation}
where $p = \frac12(p_3 + p_4)$ is the average momentum.

We need to consider this additional term at each order, however it mostly does not contribute beyond the leading term. 

With this in mind, the fully normalised contribution to the classical potential is then given by
\begin{align}\label{mainls}
\braket{\cl{M}^s} &= -\left(\frac{\kappa}{2}\right)^2\frac{e^{-K\cdot a}}{t}\left(u^2\left(1 + \frac{K\cdot a}{s}\right)^{2s} + v^2\right)\\&~~~~~~~~ - \frac{(\kappa e)^2m_Ae^{-K\cdot a}}{32\sqrt{-t}}\oint_{\Gamma} \frac{\sd y}{y} \left[ \left(1-\frac{\epsilon(1+y)^2}{2y}\right)^2\left(1+\frac{(1+y)^2}{4y}\frac{K\cdot a}{s}\right)^{2s}\right]\nonumber, 
\end{align}
where the brackets signify that we have evaluated the GEV.

With this in hand, we can now compute various pieces of the classical potential, matching to the literature where possible. 

The classical potential $V(r)$ for a gravitomagnetic system is of the form
\begin{equation}\label{key}
V(r) = m\left(\Phi(r) + \varphi(r)\textbf{a}\cdot\textbf{B}\right),
\end{equation}
where $\Phi$ is the gravitational potential and $\textbf{B}$ is gravitomagnetic field $B^i = \epsilon^{ijk}\partial_jw_k$. We note that to identify $w_i$, it will enter the momentum space potential with a factor of $p/m$. 

In order to construct the potential from the scattering amplitudes, we construct the momentum space potential as a function of transfer momentum $\textbf{q}$ and then Fourier transform to find
\begin{equation}\label{key}
V(r) = \int d^3\hat{\textbf{q}} e^{i\textbf{q}\cdot \textbf{r}}V(\textbf{q}) = \int d^3\hat{\textbf{q}} e^{i\textbf{q}\cdot \textbf{r}}\frac{\cl{M}}{4E_AE_B}.
\end{equation}

We can also construct the metric by relating its components with the potential. The standard decomposition of the metric into its component representations is given by
\begin{equation}\label{key}
h_{00} = 2\Phi,~~~~~h_{0i} = -w_i,~~~~~h_{ij} = 2\Phi\delta_{ij},
\end{equation}
where we have assumed that the scalar components are equal to one another since we are interested in the non-relativistic limit. To identify the scalar part of the metric from the potential with probe mass $m$, we can take
\begin{equation}\label{key}
\Phi = \lim_{m\longrightarrow 0, a\longrightarrow 0}\frac{1}{m}V(r).
\end{equation}
\subsection{Spin-Independent Potential}
The simplest place we can start is with the spin-independent contribution to the momentum space potential, from which we can derive the Reissner-Nordstr\"om metric. We begin by noting that the spinless limit is arrived at easily, taking the limit of $u,v\longrightarrow m_Am_B$ and $s\rightarrow 0$ in eq. \ref{mainls}, finding
\begin{equation}\label{key}
\braket{\cl{M}^0} = -\left(\frac{\kappa^2}{2}\right)\frac{m_A^2m_B^2}{t} -(\kappa e)^2\frac{m_A^2m_B}{16\sqrt{-t}},
\end{equation}
We can now compute the momentum space potential for a given spin. 
\begin{equation}\label{key}
V(\textbf{q})^{s=0} = \frac{\braket{\cl{M}^0}}{4m_Am_B} = \frac{4\pi G m_Am_B}{\textbf{q}^2} - \frac{Gm_A\pi^2\alpha}{|\textbf{q}|},
\end{equation}
where the first term is nothing more than the standard Newtownian potential in momentum space.

In position space, this is given by
\begin{equation}\label{key}
V(r) = \frac{Gm_Am_B}{r}-\frac{Gm_A\alpha}{2r^2},
\end{equation}
from which we identify a metric of the form
\begin{align}\label{key}
g_{00} &= 1-\frac{2Gm_B}{r} + \frac{G\alpha}{r^2} + \cl{O}(G^2,\alpha^2)\nonumber\\
g_{0i} &= 0,\\
g_{ij} &= \delta_{ij} -\delta_{ij}\frac{2Gm_B}{r} + \delta_{ij}\frac{G\alpha}{r^2} + \cl{O}(G^2,\alpha^2)\nonumber
\end{align}
which is precisely the Reissner-Nordstr\"om metric.

\subsection{Spin--Orbit Potential}
We now consider a non-zero spin $s$ external particle in order to extract a spin-dependent piece of the potential. While the universality of gravity dictates that the potential be the same for any spin $s$, for simplicity we choose $s = 1$. We have checked explicitly that universality of this piece of the potential holds at least up to  $s = 8$. We find that for $s = 1$ the GEV of the amplitude is
\begin{equation}
\braket{\cl{M}^1} = \kappa^2\frac{(m_Am_B)^2}{2\textbf{q}^2} - (\kappa e)^2\frac{m_A^2m_B}{16|\textbf{q}|} + \left(\kappa^2\frac{m_Am_B(m_A+m_B)}{\textbf{q}^2} - (\kappa e)^2\frac{m_A(m_A+m_B)}{16|\textbf{q}|}\right)(i\textbf{a}\cdot (\textbf{p}\times \textbf{q})),
\end{equation}
where we have taken the $\rho\rightarrow 1$ limit after identifying the relevant spin interactions. 

The first thing to note is that the first two terms are the universal spin-independent pieces, as anticipated due to the equivalence principle. The second two terms are the first-order in spin-orbit corrections. However, while this amplitude is correct, at this order we will also need to include the additional piece that comes from eq. \ref{polcontribution}. This effectively means we need to add the following term to the potential
\begin{equation}\label{key}
\frac{4\pi G m_Am_B}{\textbf{q}^2}\epsilon_3^\star\cdot\epsilon_4\Bigg|_{\text{spin}} \sim -\frac{2\pi Gm_A}{\textbf{q}^2}(i\textbf{a}\cdot (\textbf{p}\times \textbf{q})).
\end{equation}

Putting this all together, the momentum space potential is then given by
\begin{equation}\label{key}
V(\textbf{q}) = \frac{4\pi G m_Am_B}{\textbf{q}^2} - \frac{\pi^2Gm_A\alpha}{|\textbf{q}|} + \left(\frac{2\pi G(3m_A+4m_B)}{\textbf{q}^2} - \frac{\pi^2G\alpha(m_A+m_B)}{m_B|\textbf{q}|}\right)(i\textbf{a}\cdot (\textbf{p}\times \textbf{q})).
\end{equation}
Performing the Fourier transforms, we then find
\begin{equation}\label{key}
V(r) = \frac{Gm_Am_B}{r}-\frac{Gm_A\alpha}{2r^2} - \left(\frac{G(3m_A+4m_B)}{2r^3} + \frac{G\alpha(m_A+m_B)}{m_Br^4}\right)(\textbf{a}\cdot (\textbf{p}\times\textbf{r})), 
\end{equation}
from which we can identify the components of the metric
\begin{align}\label{key}
g_{00} &= 1-\frac{2Gm_B}{r} + \frac{G\alpha}{r^2} + \cl{O}(G^2,\alpha^2)\nonumber\\
g_{0i} &= \left(\frac{2Gm_B}{r^3} - \frac{G\alpha}{r^4}\right)(\textbf{a}\times\textbf{r})_i + \cl{O}(G^2,\alpha^2),\\
g_{ij} &= \delta_{ij} -\delta_{ij}\frac{2Gm_B}{r} + \delta_{ij}\frac{G\alpha}{r^2} + \cl{O}(G^2,\alpha^2)\nonumber,
\end{align}
which is the Kerr-Newman metric at order $\cl{O}(G,\alpha)$. We see then that the relation between the Reissner-Nordstr\"om metric and the Kerr-Newman metric at this order is precisely given by exposing the spin dependence of the minimally coupled three-point amplitudes of the spinning particles, specifically giving rise to the $g_{0i}$ terms in the metric. In order to sharpen this point, in the next section we will take the infinite spin limit and compute the classical impulse imparted to the probe particle.
\subsection{Infinite Spin Limit}
While we could continue to compute higher order in spin corrections, if we were so inclined, we will instead take a slightly different path in this section, and simply take the infinite spin limit.
The intrinsic angular momentum of a spin $s$ particle scales like $\braket{a^\mu} \propto s\hbar$. This means that, when considering spin, a fully consistent classical limit is only reached by taking $s\longrightarrow \infty$ as $\hbar\longrightarrow 0$ keeping $s\hbar$ (and therefore $\braket{a^\mu}$) finite \cite{Maybee:2019jus}. We now make a further identification for the variables $u$ and $v$ as being
\begin{equation}\label{key}
u = m_Am_B\gamma(1+\varv) = m_Am_Be^w,~~~~~v = m_Am_B\gamma(1-\varv) = m_Am_Be^{-w},
\end{equation}
where $w$ is the rapidity and $\gamma$ the usual Lorentz factor. Plugging this into the four-point amplitude eq. \ref{fourpt} and taking the infinite spin limit we find
\begin{align}\label{key}
\cl{I}_\infty &= \frac{gm_A^2m_B}{16\sqrt{-t}}\oint_{\Gamma} \frac{\sd y}{y} \left[\left(\cosh w - \frac{(1 + y^2)}{2y}\sinh w \right)^2e^{K\cdot a}\sum_{n = -\infty}^\infty I_n(K\cdot a)y^n\right],
\end{align}
and therefore
\begin{equation}\label{key}
\braket{\cl{I}_\infty} = \frac{gm_A^2m_B}{16\sqrt{-t}}\left[\frac12\left(2\cosh^2w - \sinh^2 w\right)I_0(K\cdot a) -2\cosh w\sinh wI_1(K\cdot a) + \frac12\sinh^2w I_2(K\cdot a)\right]
\end{equation}
where we recognise the generating function $e^{\frac{1}{2}z(y + 1/y)} = \sum I_n(z) y^n$, where $I_n$ is the modified Bessel function.


Similarly, we can do the same for eq. \ref{kerr4pt} which gives
\begin{align}\label{key}
\braket{\cl{M}_4^\infty} &= \left(\frac{\kappa}{2}\right)^2\frac{1}{t}\left(u^2e^{K\cdot a} + v^2e^{-K\cdot a}\right)\\
&= \left(\frac{\kappa}{2}\right)^2\frac{m_A^2m_B^2}{t}\left(e^{2w}e^{\textbf{q}\cdot \textbf{a}} + e^{-2w}e^{-\textbf{q}\cdot \textbf{a}}\right).
\end{align}

This allows us to cast the infinite-spin amplitude into the form
\begin{align}\label{key}
\braket{\cl{M}^\infty} = &\frac{gm_A^2m_B}{16\sqrt{-t}}\left[\frac12\left(2\cosh^2w - \sinh^2 w\right)I_0(K\cdot a) -2\cosh w\sinh wI_1(K\cdot a) + \frac12\sinh^2w I_2(K\cdot a)\right]\nn\\&~~~+ \left(\frac{\kappa}{2}\right)^2\frac{m_A^2m_B^2}{t}\left(e^{2w}e^{\textbf{q}\cdot \textbf{a}} + e^{-2w}e^{-\textbf{q}\cdot \textbf{a}}\right).
\end{align}
We now move on to compute the impulse of our scalar probe particle as a result its interaction with the spinning particle. A very careful analysis of the classical impulse in terms of scattering amplitudes was carried out in \cite{Kosower:2018adc}, however for our purposes we simply need the formula
\begin{align}\label{key}
\Delta p^\mu_1 &= \frac{1}{4m_Am_B}\int \hat{d}^4\bar{q}\hat{\delta}(\bar{q}\cdot u_1)\hat{\delta}(\bar{q}\cdot u_3)e^{-i\bar{q}\cdot b}i\bar{q}^\mu \braket{\cl{M}^\infty}.
\end{align}
The impulse is given in terms the incoming probe particle momentum $p_1 = m_Au_1$ and its colliding partner $p_3 = m_Bu_3$, and is simply a measure of the total change in momentum of particle 1 as a result of the collision.
 
The pure gravity minimally--coupled piece was computed in Ref. \cite{Arkani-Hamed:2019ymq} and found to be
\begin{align}\label{key}
\Delta p^\mu_{1,\kappa^2} &= \frac{1}{4m_Am_B}\int \hat{d}^4\bar{q}\hat{\delta}(\bar{q}\cdot u_1)\hat{\delta}(\bar{q}\cdot u_3)i\bar{q}^\mu \frac{ie^{i\bar{q}\cdot(b-i\Pi a)}}{\bar{q}^2}(\bar{q}^\mu \cosh 2w + 2i\cosh w \epsilon_{\mu\nu\rho\sigma}\bar{q}^\nu u_1^\rho u_3^\sigma).
\end{align}
In order to derive the piece of the impulse that corresponds to the charged solution, we first note a useful identity \cite{Arkani-Hamed:2019ymq}
\begin{equation}\label{key}
\sinh w \bar{q}_\mu = i\epsilon_{\mu\nu\rho\sigma}\bar{q}^\nu u_1^\rho u_3^\sigma.
\end{equation}

Defining $\overline{dq} = \hat{d}^4\bar{q}\hat{\delta}(\bar{q}\cdot u_1)\hat{\delta}(\bar{q}\cdot u_3)$, we find
\begin{align}
\Delta p^\mu_{1,(\kappa e)^2} &= \frac{gm_A^2m_B}{16}\int \frac{\overline{dq}}{|q|}e^{-i\bar{q}\cdot b}\left[\left(q^\mu + \frac{i}{2}\sinh w \zeta^\mu\right) I_0 - i2\cosh w \zeta^\mu I_1 + \frac{i}{2}\sinh w \zeta^\mu I_2\right]\nn\\
&= \frac{gm_A^2m_B}{16\pi}\int \frac{\overline{dq}}{|q|}\int_0^\pi d\theta e^{-i\bar{q}\cdot(b + ia\cos\theta)}\nn\\
&~~~~~~~~~~~~~~\times\left[q^\mu + \frac{i}{2}\sinh w \zeta^\mu - i2\cosh w \zeta^\mu \cos\theta + \frac{i}{2}\sinh w \zeta^\mu \cos 2\theta\right]\\
&= \frac{gm_A^2m_B}{16\pi}\int \frac{\overline{dq}}{|q|}\int_0^\pi d\theta e^{-i\bar{q}\cdot(b + ia\cos\theta)}\left[q^\mu + i\sinh w \zeta^\mu\cos^2\theta - i2\cosh w \zeta^\mu \cos\theta\right]\nn
\end{align}
where $\zeta^\mu \coloneqq \epsilon^{\mu\nu\rho\sigma}q_\nu u_{1\rho} u_{3\sigma}$.

The full impulse for the Kerr-Newman system, at order $\cl{O}(G,\alpha)$, is therefore given by
\begin{align}\label{key}
\Delta p^\mu_{1} &= \Re\Bigg[\int\overline{dq}\Bigg(-\frac{4\pi Gm_Am_B}{\bar{\textbf{q}}^2}(\bar{q}^\mu \cosh 2w + 2i\cosh w \epsilon_{\mu\nu\rho\sigma}\bar{q}^\nu u_1^\rho u_3^\sigma)\Bigg)e^{-i\textbf{q}\cdot (\textbf{b} +i\Pi\textbf{a})}\Bigg]\nonumber\\
&~~~~~~~~~~~+ 4\pi G \alpha m_A^2m_B\int \frac{\overline{d\textbf{q}}}{|\textbf{q}|}\int_0^\pi d\theta \left[\textbf{q}^\mu + i\sinh w \zeta^\mu\cos^2\theta - i2\cosh w \zeta^\mu \cos\theta\right]e^{-i\bar{\textbf{q}}\cdot(\textbf{b} + i\textbf{a}\cos\theta)}.
\end{align}

We see then that we can identify the shift in the Kerr-Newman solution as arising from the exponentiation of minimal coupling amplitudes, as was pointed out in the Kerr case in Ref. \cite{Arkani-Hamed:2019ymq}. We observe specifically that the impulse for Kerr-Newman is obtained when the impact factor undergoes a complex shift.

Evaluating the Fourier and Elliptical integrals as in appendix A, we then find that the impulse is
\begin{align}\label{key}
\Delta p_1^\mu &= -\frac{2Gm_Am_B}{\sinh w}\Re\left[\frac{\tilde{b}_\perp^\mu \cosh 2w + 2i\cosh w \epsilon^{\mu\nu\rho\sigma}\tilde{b}_{\perp\nu} u_{1\rho} u_{3\sigma}}{|\tilde{b}_\perp|^2}\right]\nonumber \\
&~~~+ \frac{4\pi G\alpha m_A^2m_B}{\sinh w}\Re\left[\frac{\hat{b}_\perp^\mu + i\sinh w \epsilon^{\mu\nu\rho\sigma}\hat{b}_{\perp\nu} u_{1\rho} u_{3\sigma}}{|\tilde{b}_\perp|^2} \right],
\end{align}
where we have used the relation $|\beta\gamma| = \sinh w$ and $\tilde{b}_\perp = b_\perp+ i\Pi a = \Pi(b+ia)$, and the hats indicate unit vectors.
\section{Discussion}
In this paper we have demonstrated that the leading singularity together with minimal coupling can efficiently characterize all asymptotically flat four dimensional black hole solutions at 1PN. Furthermore, we have shown that the exponentiation of minimally coupled amplitudes (in the infinite spin limit) is the on-shell avatar of the Newman-Janis algorithm that relates the Reissner-Nordst\"om and Kerr-Newman solutions. Moreover, we find that the spin-independent and spin-dependent parts of all black hole solutions factorise, reflecting the universal nature of gravity.  

In this work we have only considered a scalar probe particle, however it is almost trivial to couple a spinning particle to a charged black hole using this formalism: we simply include a spin factor for the gravitational three-point. Furthermore, giving the probe particle both spin and charge would mean the scattering of two Kerr-Newman black holes could be considered, as was done recently in the Kerr case \cite{Guevara:2018wpp,Guevara:2019fsj}. It would also be interesting to derive the all order in spin potential using the HCL \cite{Chung:2019duq}.

While we have focused on a conservative system here, the general formalism for extracting spin dependence in observables can be used for non-conservative systems \cite{Bautista:2019tdr}. One could for example consider electromagnetic or gravitational radiation being emitted by the charged/spinning particles during a scattering event and the results in this paper could be adapted easily to such a situation. It is expected that nearly all realistic black holes in the universe will be spinning, therefore these kinds of calculations would provide important theoretical predictions that could then be compared with data from both current and future gravitational wave experiments, along with their optical counterparts.

Another natural follow-up to this work is to explore higher order in $G$ black hole solutions that arise from non-minimal coupling, such as those that arise in \textit{Einsteinian Cubic Gravity} (ECG) \cite{Bueno:2016xff,Bueno:2016lrh,Hennigar:2016gkm,Emond:2019crr}. Intriguingly, no spinning solution currently exists in ECG and in principle such a solution could easily be found via the leading singularity (as was done in \cite{Emond:2019crr} for the static case). A compelling reason to carry out this study is to see whether or not deriving a solution via amplitudes will lead to a Newman-Janis type complex coordinate deformation that relates the spinning and static cases. We leave these explorations for the future.
\section*{Acknowledgements}
This research was supported by funding from the DST/NRF SARChI in Physical Cosmology.  Any opinion, finding and conclusion or recommendation expressed in this material is that of the authors and the NRF does not accept any liability in this regard.
\appendix
\section{Integral Transforms}
We collect here some useful integral transforms that were used throughout this paper.
\begin{align}\label{FTs}
\int \frac{\sd^3\mathbf{q}}{(2\pi)^3}e^{i\mathbf{q}\cdot \mathbf{r}}|\mathbf{q}|^n &=
\frac{(n+1)!}{2\pi^2r^{3+n}}\sin\left(\frac{3\pi n}{2}\right)\\
\int \frac{\sd^3\mathbf{q}}{(2\pi)^3}e^{i\mathbf{q}\cdot \mathbf{r}}\frac{q_j}{|\mathbf{q}|} &= \frac{ir_j}{\pi^2r^4}\\
\int \frac{\sd^3\mathbf{q}}{(2\pi)^3}e^{i\mathbf{q}\cdot \mathbf{r}}\frac{q_j}{\mathbf{q}^2} &= \frac{ir_j}{4\pi r^3}\\
\int \frac{\sd^3\mathbf{q}}{(2\pi)^3}e^{i\mathbf{q}\cdot \mathbf{r}}\frac{q_jq_k}{\textbf{q}^2} &= \frac{1}{3}\delta_{jk}\delta(\textbf{r}) + \frac{1}{4\pi r^3}(\delta_{jk} - 3\frac{r_jr_k}{r^2})\\
\int \frac{\sd^3\mathbf{q}}{(2\pi)^3}e^{i\mathbf{q}\cdot \mathbf{r}}\frac{q_jq_k}{|\textbf{q}|} &= \frac{1}{\pi^2 r^4}(\delta_{jk} - 4\frac{r_jr_k}{r^2})\\
\int_{-\pi}^\pi d\theta~e^{-i|p||r|\cos\theta}\cos\theta &= -2\pi i J_1(|p||r|),
\end{align}
where $J_1$ is a Bessel function of the first kind.

The Hankel transform of $r^n$ is given by
\begin{equation}\label{key}
\cl{H}_\nu[r^n] = \int_0^\infty r^{n+1} J_\nu(kr) = \frac{2^{n+1}}{k^{n+2}}\frac{\Gamma(\frac12(2 + \nu + n))}{\Gamma(\frac12(\nu - n))}  
\end{equation}
\subsection{Impulse Fourier Transform}
To compute the Fourier transform needed for the Kerr-Newman impulse, we need to evaluate the following integrals
\begin{align}\label{key}
\mathfrak{F}\left[\frac{q^\mu}{|q|}\right] &= \int \hat{d}^4\bar{q}\hat{\delta}(\bar{q}\cdot u_1)\hat{\delta}(\bar{q}\cdot u_3) \frac{\bar{q}^\mu}{|\bar{q}|}e^{-i\bar{q}\cdot \tilde{b}},~~~~~\tilde{b} = b + i\Pi a\\
\mathfrak{F}\left[\frac{q^\mu}{\bar{q}^2}\right] &= \int \hat{d}^4\bar{q}\hat{\delta}(\bar{q}\cdot u_1)\hat{\delta}(\bar{q}\cdot u_3) \frac{\bar{q}^\mu}{\bar{q}^2}e^{-i\bar{q}\cdot \tilde{b}}.
\end{align}
We can evaluate these following Ref. \cite{Kosower:2018adc} by working in the rest frame of particle 1, meaning we take
\begin{equation}\label{key}
u_1 = (1,0,0,0),~~~~~ u_3 = (\gamma,0,0,\gamma\beta).
\end{equation}
In this frame, we find that the delta functions enforce $\bar{q}^0 = \bar{q}^3 = 0$ and that the integral reduces to a two dimensional integral over the components orthogonal to $u_1$ and $u_3$, e.g.
\begin{align}\label{key}
\int \hat{d}^4\bar{q}\hat{\delta}(\bar{q}\cdot u_1)\hat{\delta}(\bar{q}\cdot u_3) \frac{\bar{q}^\mu}{|\bar{q}|}e^{-i\bar{q}\cdot \tilde{b}} &= \int \hat{d}^4\bar{q}\hat{\delta}(\bar{q}^0)\hat{\delta}(\gamma \bar{q}^1 - \beta\gamma\bar{q}^3) \frac{\bar{q}^\mu}{|\bar{q}|}e^{-i\bar{q}\cdot \tilde{b}}\nonumber \\
&= \frac{1}{4\pi^2 |\beta\gamma|}\int \hat{d}^2\bar{\textbf{q}}_\perp e^{-i\bar{\textbf{q}}_\perp\cdot\tilde{\textbf{b}}}\frac{\bar{q}^\mu}{|\bar{\textbf{q}}_\perp|}
\end{align}

Evaluating these (in polar coordinates) we find
\begin{align}
\mathfrak{F}\left[\frac{q^\mu}{|q|}\right] &= -\frac{1}{4\pi^2|\beta\gamma|}\int \hat{d}^2\bar{\textbf{q}}_\perp e^{-i\bar{\textbf{q}}_\perp\cdot\tilde{\textbf{b}}}\frac{\bar{q}^\mu}{|\bar{\textbf{q}}_\perp|}\\
&= -\frac{1}{4\pi^2|\beta\gamma|}\int_0^\infty d\chi \int_{-\pi}^\pi d\theta e^{-i\chi|\tilde{\textbf{b}}|\cos\theta}\bar{q}^\mu\\
&= -\frac{i}{2\pi|\beta\gamma|}\int_0^\infty d\chi~\chi J_1(\chi|\tilde{\textbf{b}}|) \hat{\textbf{b}}\\
&= -\frac{i}{2\pi|\beta\gamma|}\cl{H}_1[1]\hat{\textbf{b}}\\
&= -\frac{i}{2\pi|\beta\gamma|}\frac{\hat{\textbf{b}}}{|\tilde{\textbf{b}}|^2}\\
&= \frac{i}{2\pi|\beta\gamma|}\frac{b^\mu}{|\tilde{b}|^3},
\end{align}
and
\begin{align}
\mathfrak{F}\left[\frac{q^\mu}{\bar{q}^2}\right] &= -\frac{1}{4\pi^2}\int \hat{d}^2\bar{\textbf{q}}_\perp e^{-i\bar{q}_\perp\cdot\tilde{b}}\frac{\bar{q}^\mu}{\bar{\textbf{q}}_\perp^2}\\
&= -\frac{1}{4\pi^2|\beta\gamma|}\int_0^\infty d\chi \int_{-\pi}^\pi d\theta e^{-i\chi|\tilde{b}|\cos\theta}\frac{\bar{q}^\mu}{\chi}\\
&=  -\frac{i}{2\pi|\beta\gamma|}\int_0^\infty d\chi~J_1(\chi|\tilde{b}|) \hat{b}\\
&= -\frac{i}{2\pi|\beta\gamma|}\cl{H}_1[\chi^{-1}]\hat{b}\\
&= -\frac{i}{2\pi|\beta\gamma|}\frac{\hat{b}}{|\tilde{b}|}\\
&= \frac{i}{2\pi|\beta\gamma|}\frac{b^\mu}{|\tilde{b}|^2}.
\end{align}
\subsection{Elliptical Integrals}
After Fourier transforming the impulse, we are left with the following integral to evaluate
\begin{equation}\label{key}
\int_{0}^\pi d\theta \frac{1}{2\pi\sinh w|b_\perp + ia\cos\theta|^3}\left[b_\perp^\mu + i\sinh w\zeta_\perp^\mu\cos^2\theta - 2i\cosh w\zeta_\perp^\mu\cos\theta\right].
\end{equation}
The $\theta$ dependence resides in the class of elliptical integrals
\begin{equation}\label{key}
L^n = \int_0^\pi d\theta \frac{\cos^n\theta}{|b_\perp + ia\cos\theta|^3}.
\end{equation}
To compute this, we make the substitution $u = \cos\theta$ to find
\begin{equation}\label{key}
L^n = \int_{-1}^1 du \frac{u^n}{\sqrt{1-u^2}(b_\perp^2 + a^2 u^2)^{3/2}},
\end{equation}
which is an elliptical integral with well known solutions.
Computing this for large impact parameter (i.e. $b\gg a$), we find the asymptotic forms of the integrals are
\begin{equation}\label{key}
L^0 = L^2 = \frac{\pi}{|b_\perp||b_\perp + ia|^2} + \cl{O}\left(\frac{a^2}{b^2}\right),~~~~~L^1 = 0.
\end{equation}
Plugging this in, we then find
\begin{align}\label{key}
\Delta p^\mu &= \frac{gm_A^2m_B}{32\pi^2\sinh w}\left(\frac{b_\perp^\mu + i\sinh w \zeta_\perp^\mu}{|b_\perp||b_\perp + ia|^2} \right)\\
&= \frac{4\pi G\alpha m_A^2m_B}{\sinh w}\left(\frac{\hat{b}_\perp^\mu + i\sinh w \hat{\zeta}_\perp^\mu}{|b_\perp + ia|^2} \right).
\end{align}
\bibliographystyle{JHEP}
\bibliography{mainbib}

\providecommand{\href}[2]{#2}\begingroup\raggedright\begin{thebibliography}{10}

\bibitem{Arkani-Hamed:2019ymq}
N.~Arkani-Hamed, Y.-t. Huang and D.~O'Connell, \emph{{Kerr Black Holes as
  Elementary Particles}},  \href{http://arxiv.org/abs/1906.10100}{{\tt
  1906.10100}}.

\bibitem{Thirring:1961dc}
W.~E. Thirring, \emph{{An alternative approach to the theory of gravitation}},
  \href{http://dx.doi.org/10.1016/0003-4916(61)90182-8}{\emph{Annals Phys.}
  {\bf 16} (1961) 96--117}.

\bibitem{Iwasaki:1971vb}
Y.~Iwasaki, \emph{{Quantum theory of gravitation vs. classical theory. -
  fourth-order potential}},
  \href{http://dx.doi.org/10.1143/PTP.46.1587}{\emph{Prog. Theor. Phys.} {\bf
  46} (1971) 1587--1609}.

\bibitem{Duff:1973zz}
M.~J. Duff, \emph{{Quantum Tree Graphs and the Schwarzschild Solution}},
  \href{http://dx.doi.org/10.1103/PhysRevD.7.2317}{\emph{Phys. Rev.} {\bf D7}
  (1973) 2317--2326}.

\bibitem{Bern:2008qj}
Z.~Bern, J.~J.~M. Carrasco and H.~Johansson, \emph{{New Relations for
  Gauge-Theory Amplitudes}},
  \href{http://dx.doi.org/10.1103/PhysRevD.78.085011}{\emph{Phys. Rev.} {\bf
  D78} (2008) 085011}, [\href{http://arxiv.org/abs/0805.3993}{{\tt
  0805.3993}}].

\bibitem{Bern:2010ue}
Z.~Bern, J.~J.~M. Carrasco and H.~Johansson, \emph{{Perturbative Quantum
  Gravity as a Double Copy of Gauge Theory}},
  \href{http://dx.doi.org/10.1103/PhysRevLett.105.061602}{\emph{Phys. Rev.
  Lett.} {\bf 105} (2010) 061602}, [\href{http://arxiv.org/abs/1004.0476}{{\tt
  1004.0476}}].

\bibitem{Bern:2010yg}
Z.~Bern, T.~Dennen, Y.-t. Huang and M.~Kiermaier, \emph{{Gravity as the Square
  of Gauge Theory}},
  \href{http://dx.doi.org/10.1103/PhysRevD.82.065003}{\emph{Phys. Rev.} {\bf
  D82} (2010) 065003}, [\href{http://arxiv.org/abs/1004.0693}{{\tt
  1004.0693}}].

\bibitem{Johansson:2019dnu}
H.~Johansson and A.~Ochirov, \emph{{Double copy for massive quantum particles
  with spin}}, \href{http://dx.doi.org/10.1007/JHEP09(2019)040}{\emph{JHEP}
  {\bf 09} (2019) 040}, [\href{http://arxiv.org/abs/1906.12292}{{\tt
  1906.12292}}].

\bibitem{Bautista:2019evw}
Y.~F. Bautista and A.~Guevara, \emph{{On the Double Copy for Spinning Matter}},
   \href{http://arxiv.org/abs/1908.11349}{{\tt 1908.11349}}.

\bibitem{Guevara:2017csg}
A.~Guevara, \emph{{Holomorphic Classical Limit for Spin Effects in
  Gravitational and Electromagnetic Scattering}},
  \href{http://dx.doi.org/10.1007/JHEP04(2019)033}{\emph{JHEP} {\bf 04} (2019)
  033}, [\href{http://arxiv.org/abs/1706.02314}{{\tt 1706.02314}}].

\bibitem{Kosower:2018adc}
D.~A. Kosower, B.~Maybee and D.~O'Connell, \emph{{Amplitudes, Observables, and
  Classical Scattering}},
  \href{http://dx.doi.org/10.1007/JHEP02(2019)137}{\emph{JHEP} {\bf 02} (2019)
  137}, [\href{http://arxiv.org/abs/1811.10950}{{\tt 1811.10950}}].

\bibitem{Maybee:2019jus}
B.~Maybee, D.~O'Connell and J.~Vines, \emph{{Observables and amplitudes for
  spinning particles and black holes}},
  \href{http://arxiv.org/abs/1906.09260}{{\tt 1906.09260}}.

\bibitem{Neill:2013wsa}
D.~Neill and I.~Z. Rothstein, \emph{{Classical Space-Times from the S Matrix}},
  \href{http://dx.doi.org/10.1016/j.nuclphysb.2013.09.007}{\emph{Nucl. Phys.}
  {\bf B877} (2013) 177--189}, [\href{http://arxiv.org/abs/1304.7263}{{\tt
  1304.7263}}].

\bibitem{Luna:2016hge}
A.~Luna, R.~Monteiro, I.~Nicholson, A.~Ochirov, D.~O'Connell, N.~Westerberg
  et~al., \emph{{Perturbative spacetimes from Yang-Mills theory}},
  \href{http://dx.doi.org/10.1007/JHEP04(2017)069}{\emph{JHEP} {\bf 04} (2017)
  069}, [\href{http://arxiv.org/abs/1611.07508}{{\tt 1611.07508}}].

\bibitem{Monteiro:2014cda}
R.~Monteiro, D.~O'Connell and C.~D. White, \emph{{Black holes and the double
  copy}}, \href{http://dx.doi.org/10.1007/JHEP12(2014)056}{\emph{JHEP} {\bf 12}
  (2014) 056}, [\href{http://arxiv.org/abs/1410.0239}{{\tt 1410.0239}}].

\bibitem{Luna:2015paa}
A.~Luna, R.~Monteiro, D.~O'Connell and C.~D. White, \emph{{The classical double
  copy for Taub-NUT spacetime}},
  \href{http://dx.doi.org/10.1016/j.physletb.2015.09.021}{\emph{Phys. Lett.}
  {\bf B750} (2015) 272--277}, [\href{http://arxiv.org/abs/1507.01869}{{\tt
  1507.01869}}].

\bibitem{Carrillo-Gonzalez:2017iyj}
M.~Carrillo-Gonz\'alez, R.~Penco and M.~Trodden, \emph{{The classical double
  copy in maximally symmetric spacetimes}},
  \href{http://dx.doi.org/10.1007/JHEP04(2018)028}{\emph{JHEP} {\bf 04} (2018)
  028}, [\href{http://arxiv.org/abs/1711.01296}{{\tt 1711.01296}}].

\bibitem{Luna:2018dpt}
A.~Luna, R.~Monteiro, I.~Nicholson and D.~O'Connell, \emph{{Type D Spacetimes
  and the Weyl Double Copy}},
  \href{http://dx.doi.org/10.1088/1361-6382/ab03e6}{\emph{Class. Quant. Grav.}
  {\bf 36} (2019) 065003}, [\href{http://arxiv.org/abs/1810.08183}{{\tt
  1810.08183}}].

\bibitem{CarrilloGonzalez:2019gof}
M.~Carrillo~Gonz\'alez, B.~Melcher, K.~Ratliff, S.~Watson and C.~D. White,
  \emph{{The classical double copy in three spacetime dimensions}},
  \href{http://dx.doi.org/10.1007/JHEP07(2019)167}{\emph{JHEP} {\bf 07} (2019)
  167}, [\href{http://arxiv.org/abs/1904.11001}{{\tt 1904.11001}}].

\bibitem{LopesCardoso:2018xes}
G.~Lopes~Cardoso, G.~Inverso, S.~Nagy and S.~Nampuri, \emph{{Comments on the
  double copy construction for gravitational theories}},
  \href{http://dx.doi.org/10.22323/1.318.0177}{\emph{PoS} {\bf CORFU2017}
  (2018) 177}, [\href{http://arxiv.org/abs/1803.07670}{{\tt 1803.07670}}].

\bibitem{Goldberger:2016iau}
W.~D. Goldberger and A.~K. Ridgway, \emph{{Radiation and the classical double
  copy for color charges}},
  \href{http://dx.doi.org/10.1103/PhysRevD.95.125010}{\emph{Phys. Rev.} {\bf
  D95} (2017) 125010}, [\href{http://arxiv.org/abs/1611.03493}{{\tt
  1611.03493}}].

\bibitem{Luna:2017dtq}
A.~Luna, I.~Nicholson, D.~O'Connell and C.~D. White, \emph{{Inelastic Black
  Hole Scattering from Charged Scalar Amplitudes}},
  \href{http://dx.doi.org/10.1007/JHEP03(2018)044}{\emph{JHEP} {\bf 03} (2018)
  044}, [\href{http://arxiv.org/abs/1711.03901}{{\tt 1711.03901}}].

\bibitem{Bern:2019crd}
Z.~Bern, C.~Cheung, R.~Roiban, C.-H. Shen, M.~P. Solon and M.~Zeng,
  \emph{{Black Hole Binary Dynamics from the Double Copy and Effective
  Theory}},  \href{http://arxiv.org/abs/1908.01493}{{\tt 1908.01493}}.

\bibitem{Bautista:2019tdr}
Y.~F. Bautista and A.~Guevara, \emph{{From Scattering Amplitudes to Classical
  Physics: Universality, Double Copy and Soft Theorems}},
  \href{http://arxiv.org/abs/1903.12419}{{\tt 1903.12419}}.

\bibitem{Arkani-Hamed:2017jhn}
N.~Arkani-Hamed, T.-C. Huang and Y.-t. Huang, \emph{{Scattering Amplitudes For
  All Masses and Spins}},  \href{http://arxiv.org/abs/1709.04891}{{\tt
  1709.04891}}.

\bibitem{Vaidya:2014kza}
V.~Vaidya, \emph{{Gravitational spin Hamiltonians from the S matrix}},
  \href{http://dx.doi.org/10.1103/PhysRevD.91.024017}{\emph{Phys. Rev.} {\bf
  D91} (2015) 024017}, [\href{http://arxiv.org/abs/1410.5348}{{\tt
  1410.5348}}].

\bibitem{Chung:2018kqs}
M.-Z. Chung, Y.-T. Huang, J.-W. Kim and S.~Lee, \emph{{The simplest massive
  S-matrix: from minimal coupling to Black Holes}},
  \href{http://dx.doi.org/10.1007/JHEP04(2019)156}{\emph{JHEP} {\bf 04} (2019)
  156}, [\href{http://arxiv.org/abs/1812.08752}{{\tt 1812.08752}}].

\bibitem{Guevara:2018wpp}
A.~Guevara, A.~Ochirov and J.~Vines, \emph{{Scattering of Spinning Black Holes
  from Exponentiated Soft Factors}},
  \href{http://arxiv.org/abs/1812.06895}{{\tt 1812.06895}}.

\bibitem{Guevara:2019fsj}
A.~Guevara, A.~Ochirov and J.~Vines, \emph{{Black-hole scattering with general
  spin directions from minimal-coupling amplitudes}},
  \href{http://arxiv.org/abs/1906.10071}{{\tt 1906.10071}}.

\bibitem{Porto:2005ac}
R.~A. Porto, \emph{{Post-Newtonian corrections to the motion of spinning bodies
  in NRGR}}, \href{http://dx.doi.org/10.1103/PhysRevD.73.104031}{\emph{Phys.
  Rev.} {\bf D73} (2006) 104031},
  [\href{http://arxiv.org/abs/gr-qc/0511061}{{\tt gr-qc/0511061}}].

\bibitem{Levi:2010zu}
M.~Levi, \emph{{Next to Leading Order gravitational Spin-Orbit coupling in an
  Effective Field Theory approach}},
  \href{http://dx.doi.org/10.1103/PhysRevD.82.104004}{\emph{Phys. Rev.} {\bf
  D82} (2010) 104004}, [\href{http://arxiv.org/abs/1006.4139}{{\tt
  1006.4139}}].

\bibitem{Levi:2014gsa}
M.~Levi and J.~Steinhoff, \emph{{Leading order finite size effects with spins
  for inspiralling compact binaries}},
  \href{http://dx.doi.org/10.1007/JHEP06(2015)059}{\emph{JHEP} {\bf 06} (2015)
  059}, [\href{http://arxiv.org/abs/1410.2601}{{\tt 1410.2601}}].

\bibitem{Levi:2015msa}
M.~Levi and J.~Steinhoff, \emph{{Spinning gravitating objects in the effective
  field theory in the post-Newtonian scheme}},
  \href{http://dx.doi.org/10.1007/JHEP09(2015)219}{\emph{JHEP} {\bf 09} (2015)
  219}, [\href{http://arxiv.org/abs/1501.04956}{{\tt 1501.04956}}].

\bibitem{Damour:2016gwp}
T.~Damour, \emph{{Gravitational scattering, post-Minkowskian approximation and
  Effective One-Body theory}},
  \href{http://dx.doi.org/10.1103/PhysRevD.94.104015}{\emph{Phys. Rev.} {\bf
  D94} (2016) 104015}, [\href{http://arxiv.org/abs/1609.00354}{{\tt
  1609.00354}}].

\bibitem{Vines:2016qwa}
J.~Vines and J.~Steinhoff, \emph{{Spin-multipole effects in binary black holes
  and the test-body limit}},
  \href{http://dx.doi.org/10.1103/PhysRevD.97.064010}{\emph{Phys. Rev.} {\bf
  D97} (2018) 064010}, [\href{http://arxiv.org/abs/1606.08832}{{\tt
  1606.08832}}].

\bibitem{Vines:2017hyw}
J.~Vines, \emph{{Scattering of two spinning black holes in post-Minkowskian
  gravity, to all orders in spin, and effective-one-body mappings}},
  \href{http://dx.doi.org/10.1088/1361-6382/aaa3a8}{\emph{Class. Quant. Grav.}
  {\bf 35} (2018) 084002}, [\href{http://arxiv.org/abs/1709.06016}{{\tt
  1709.06016}}].

\bibitem{Vines:2018gqi}
J.~Vines, J.~Steinhoff and A.~Buonanno, \emph{{Spinning-black-hole scattering
  and the test-black-hole limit at second post-Minkowskian order}},
  \href{http://dx.doi.org/10.1103/PhysRevD.99.064054}{\emph{Phys. Rev.} {\bf
  D99} (2019) 064054}, [\href{http://arxiv.org/abs/1812.00956}{{\tt
  1812.00956}}].

\bibitem{Levi:2018nxp}
M.~Levi, \emph{{Effective Field Theories of Post-Newtonian Gravity: A
  comprehensive review}},  \href{http://arxiv.org/abs/1807.01699}{{\tt
  1807.01699}}.

\bibitem{Holstein:2008sy}
B.~R. Holstein and A.~Ross, \emph{{Long Distance Effects in Mixed
  Electromagnetic-Gravitational Scattering}},
  \href{http://arxiv.org/abs/0802.0717}{{\tt 0802.0717}}.

\bibitem{Holstein:2008sx}
B.~R. Holstein and A.~Ross, \emph{{Spin Effects in Long Range Gravitational
  Scattering}},  \href{http://arxiv.org/abs/0802.0716}{{\tt 0802.0716}}.

\bibitem{Bjerrum-Bohr:2013bxa}
N.~E.~J. Bjerrum-Bohr, J.~F. Donoghue and P.~Vanhove, \emph{{On-shell
  Techniques and Universal Results in Quantum Gravity}},
  \href{http://dx.doi.org/10.1007/JHEP02(2014)111}{\emph{JHEP} {\bf 02} (2014)
  111}, [\href{http://arxiv.org/abs/1309.0804}{{\tt 1309.0804}}].

\bibitem{Cheung:2018wkq}
C.~Cheung, I.~Z. Rothstein and M.~P. Solon, \emph{{From Scattering Amplitudes
  to Classical Potentials in the Post-Minkowskian Expansion}},
  \href{http://dx.doi.org/10.1103/PhysRevLett.121.251101}{\emph{Phys. Rev.
  Lett.} {\bf 121} (2018) 251101}, [\href{http://arxiv.org/abs/1808.02489}{{\tt
  1808.02489}}].

\bibitem{Bjerrum-Bohr:2018xdl}
N.~E.~J. Bjerrum-Bohr, P.~H. Damgaard, G.~Festuccia, L.~Plante and P.~Vanhove,
  \emph{{General Relativity from Scattering Amplitudes}},
  \href{http://dx.doi.org/10.1103/PhysRevLett.121.171601}{\emph{Phys. Rev.
  Lett.} {\bf 121} (2018) 171601}, [\href{http://arxiv.org/abs/1806.04920}{{\tt
  1806.04920}}].

\bibitem{Carballo-Rubio:2018bmu}
R.~Carballo-Rubio, F.~Di~Filippo and N.~Moynihan, \emph{{Taming
  higher-derivative interactions and bootstrapping gravity with soft
  theorems}},  \href{http://arxiv.org/abs/1811.08192}{{\tt 1811.08192}}.

\bibitem{Cristofoli:2019neg}
A.~Cristofoli, N.~E.~J. Bjerrum-Bohr, P.~H. Damgaard and P.~Vanhove, \emph{{On
  Post-Minkowskian Hamiltonians in General Relativity}},
  \href{http://arxiv.org/abs/1906.01579}{{\tt 1906.01579}}.

\bibitem{Bern:2019nnu}
Z.~Bern, C.~Cheung, R.~Roiban, C.-H. Shen, M.~P. Solon and M.~Zeng,
  \emph{{Scattering Amplitudes and the Conservative Hamiltonian for Binary
  Systems at Third Post-Minkowskian Order}},
  \href{http://dx.doi.org/10.1103/PhysRevLett.122.201603}{\emph{Phys. Rev.
  Lett.} {\bf 122} (2019) 201603}, [\href{http://arxiv.org/abs/1901.04424}{{\tt
  1901.04424}}].

\bibitem{Mazur:1982db}
P.~O. Mazur, \emph{{PROOF OF UNIQUENESS OF THE KERR-NEWMAN BLACK HOLE
  SOLUTION}}, \href{http://dx.doi.org/10.1088/0305-4470/15/10/021}{\emph{J.
  Phys.} {\bf A15} (1982) 3173--3180}.

\bibitem{Donoghue:2001qc}
J.~F. Donoghue, B.~R. Holstein, B.~Garbrecht and T.~Konstandin, \emph{{Quantum
  corrections to the Reissner-Nordstrom and Kerr-Newman metrics}},
  \href{http://dx.doi.org/10.1016/S0370-2693(02)01246-7,
  10.1016/j.physletb.2005.03.018}{\emph{Phys. Lett.} {\bf B529} (2002)
  132--142}, [\href{http://arxiv.org/abs/hep-th/0112237}{{\tt
  hep-th/0112237}}].

\bibitem{BjerrumBohr:2002ks}
N.~E.~J. Bjerrum-Bohr, J.~F. Donoghue and B.~R. Holstein, \emph{{Quantum
  corrections to the Schwarzschild and Kerr metrics}},
  \href{http://dx.doi.org/10.1103/PhysRevD.68.084005,
  10.1103/PhysRevD.71.069904}{\emph{Phys. Rev.} {\bf D68} (2003) 084005},
  [\href{http://arxiv.org/abs/hep-th/0211071}{{\tt hep-th/0211071}}].

\bibitem{Emond:2019crr}
W.~T. Emond and N.~Moynihan, \emph{{Scattering Amplitudes, Black Holes and
  Leading Singularities in Cubic Theories of Gravity}},
  \href{http://arxiv.org/abs/1905.08213}{{\tt 1905.08213}}.

\bibitem{Chung:2019duq}
M.-Z. Chung, Y.-t. Huang and J.-W. Kim, \emph{{From quantized spins to rotating
  black holes}},  \href{http://arxiv.org/abs/1908.08463}{{\tt 1908.08463}}.

\bibitem{Newman:1965tw}
E.~T. Newman and A.~I. Janis, \emph{{Note on the Kerr spinning particle
  metric}}, \href{http://dx.doi.org/10.1063/1.1704350}{\emph{J. Math. Phys.}
  {\bf 6} (1965) 915--917}.

\bibitem{Newman:1965my}
E.~T. Newman, R.~Couch, K.~Chinnapared, A.~Exton, A.~Prakash and R.~Torrence,
  \emph{{Metric of a Rotating, Charged Mass}},
  \href{http://dx.doi.org/10.1063/1.1704351}{\emph{J. Math. Phys.} {\bf 6}
  (1965) 918--919}.

\bibitem{Cachazo:2017jef}
F.~Cachazo and A.~Guevara, \emph{{Leading Singularities and Classical
  Gravitational Scattering}},  \href{http://arxiv.org/abs/1705.10262}{{\tt
  1705.10262}}.

\bibitem{Burger:2017yod}
D.~J. Burger, R.~Carballo-Rubio, N.~Moynihan, J.~Murugan and A.~Weltman,
  \emph{{Amplitudes for astrophysicists: known knowns}},
  \href{http://dx.doi.org/10.1007/s10714-018-2475-0}{\emph{Gen. Rel. Grav.}
  {\bf 50} (2018) 156}, [\href{http://arxiv.org/abs/1704.05067}{{\tt
  1704.05067}}].

\bibitem{Bueno:2016xff}
P.~Bueno and P.~A. Cano, \emph{{Einsteinian cubic gravity}},
  \href{http://dx.doi.org/10.1103/PhysRevD.94.104005}{\emph{Phys. Rev.} {\bf
  D94} (2016) 104005}, [\href{http://arxiv.org/abs/1607.06463}{{\tt
  1607.06463}}].

\bibitem{Bueno:2016lrh}
P.~Bueno and P.~A. Cano, \emph{{Four-dimensional black holes in Einsteinian
  cubic gravity}},
  \href{http://dx.doi.org/10.1103/PhysRevD.94.124051}{\emph{Phys. Rev.} {\bf
  D94} (2016) 124051}, [\href{http://arxiv.org/abs/1610.08019}{{\tt
  1610.08019}}].

\bibitem{Hennigar:2016gkm}
R.~A. Hennigar and R.~B. Mann, \emph{{Black holes in Einsteinian cubic
  gravity}}, \href{http://dx.doi.org/10.1103/PhysRevD.95.064055}{\emph{Phys.
  Rev.} {\bf D95} (2017) 064055}, [\href{http://arxiv.org/abs/1610.06675}{{\tt
  1610.06675}}].

\end{thebibliography}\endgroup
\end{document}